O. de Melo[*], A. Domínguez, K. Gutiérrez Z-B, G. Contreras-Puente[1], S. Gallardo-Hernández[2], A. Escobosa[2], J. C. González[3], R. Paniago[3], J. Ferraz Dias[4], M. Behar[4]

Physics Faculty, University of Havana, 10400 La Habana, Cuba

[1]Escuela Superior de Fisica y Matemáticas, IPN, U. P. ALM. 07738 México D.F.

[2]Sección de Electrónica del Estado Sólido, Departamento de Ingeniería Eléctrica, CINVESTAV, IPN, Mexico

[3]Departamento de Fisica, Universidade Federal de Minas Gerais, Belo Horizonte, MG 30123-970, Brazil

[4]Ion Implantation Laboratory, Physics Institute, Federal University of Rio Grande do Sul, CP 15051, CEP 91501-970, Porto Alegre, RS, Brazil

Abstract. Graded composition $Cd_xZn_{1-x}Te$ films were prepared by growing several alternate layers of CdTe and ZnTe by the Isothermal Close Space Sublimation technique. The thickness of both kinds of layers was modified along the structure to produce an increase of the average concentration of CdTe towards the surface of the films. Due to Zn/Cd inter-diffusion during the growth process the sequence of layers was converted into a relatively smooth graded composition film. According to X- ray diffraction characterization the layers grew coherently with the (100) oriented GaAs substrates although they showed a relatively high mosaicity. $\theta-2\theta$ plots show very wide diffraction peaks as expected from variable composition samples. The band gap grading effect in light absorption was also verified through transmission measurements, using transparent substrates. Graded composition profiles of the thin films were confirmed by x-ray photoelectron and secondary ion mass spectroscopies. Moreover, quantitative Cd, Zn and Te profiles were obtained by the analysis of Rutherford backscattering spectra of the samples. This analysis showed a CdTe molar fraction profile ranging from approximately x = 0.8 at the surface of the sample and x = 0.35 at the interface with the substrate. The possibility of growing graded composition layers using a simple and cost-effective procedure was demonstrated. This could be interesting in technological applications like $Cd_xZn_{1-x}Te$ layers of variable composition in CdS/CdTe solar cells.

[*]Corresponding author (e-mail: odemelo@gmail.com). Present address: Departamento de Fisica, Universidade Federal de Minas Gerais, Belo Horizonte, MG 30123-970, Brazil

1. Introduction.

Thin film semiconductor alloys with graded composition (GC) result in spatial variation of several properties such as the refractive index, band gap, effective mass, and light absorption coefficient among others. This fact has encouraged several applications in practical devices and in particular in solar cells technology. In fact, $Cd_xZn_{1-x}Te$ and $Cu(In_xGa_{1-x})Se_2$ with different composition profiles have been proposed as absorbers in order to increase the efficiency of CdTe and $Cu(In_xGa_{1-x})Se_2$ based thin film solar cells [1,2]. In these cases, the band gap variation leads to a quasi-electric field that enhances the drift-diffusion length and reduce the back surface recombination velocity. Also, $Cd_xZn_{1-x}Te$ with constant Cd composition and band gap ranging between 1.6 and 1.7 eV (x ranging around 0.2 - 0.45) has been proposed for top absorber in tandem solar cells [3, 4].

Other applications of GC films include variable refractive index profiles used in optical filters or antireflection coatings [5,6] and strain relief by accommodation of lattice misfit in heterostructures [7]. In particular, CdTe epitaxial growth on GaAs substrates has been proved to be a difficult task due to the large mismatch between both materials. Because of that ZnTe buffer layers are used to favor the epitaxial growth and to enhance the crystal quality of CdTe layers [8,9].

Separate confinement heterostructure lasers use refractive index grading (induced by GC) in order to enhance carrier trapping efficiency, trapping time and carrier confinement [10]. GC structures have been also used in thermoelectric materials to enhance the figure of merit of the devices locating the most suitable material composition in the temperature profile along the structure [11]. Finally, ohmic contacts preparation [12], increased resistance of surfaces to damage or contact deformation [13] are also process beneficiaries of GC materials.

Several techniques have been employed to prepare GC layers. In relation with solar cells applications, CdZnTe/CdTe heterostructures have been made using co-deposition of CdTe and ZnTe compounds or sputtering using $Cd_{0.6}Zn_{0.4}Te$ targets. On the other hand, CdTe/ZnTe bilayers have been used to obtain both uniform or graded profiles in $Cd_xZn_{1-x}Te$ films using 300 nm thick CdTe layers and 300 or 500 nm thick ZnTe layers grown by RF sputtering [14].

Electron Cyclotron Resonance - Plasma Enhanced Chemical Vapor Deposition (ECR-PECVD) [15] was used to grow $SiO_xN_y$ alloys with different composition profiles while Molecular Beam Epitaxy (MBE) has been frequently used for growing high quality GC layers among them $ZnTe_xSe_{1-x}$ [12] and $Pb_xEu_{1-x}Se$ [16].

In this paper we demonstrate the use of the Isothermal Close Space Sublimation (ICSS), a cost - effective technique, to prepare $Cd_xZn_{1-x}Te$ alloys with GC. This physical vapor deposition technique, in which the substrate is exposed sequentially to the elemental sources, has been previously employed for growing epitaxial as well as polycrystalline ZnTe, CdTe and $Cd_xZn_{1-x}Te$ with constant composition [17,18]. This technique revealed a regulated growth rate determined by the multilayer adsorption of the vapor elements in every exposure, the growth rate being controlled by both the exposure and the purge time.

Samples grown on GaAs (100) as well as transparent substrates (glass/ITO) were characterized by x-ray diffraction (XRD), x-ray photoelectron (XPS) and optical transmission and secondary ion mass (SIMS) spectroscopies, and Rutherford Backscattering (RBS) Spectrometry.

2. Experimental.

2.1. Sample preparation

ICSS procedure has been described elsewhere [17]. In this technique the growth of the films proceeds by cyclic exposure of the substrate to the elemental sources (Cd and Te for CdTe layers, Zn and Te for ZnTe layers). These sources are heated in a graphite crucible inside a quartz reactor with flowing gas or vacuum atmosphere.

To produce the GC samples, thin layers of CdTe and ZnTe were grown alternately in such a way that the thickness of both layers (controlled by the number of cycles) was modified along the structure to produce an increase of the average concentration of CdTe towards the surface of the sample. Different GC samples were studied in the present work. Sample A was grown with the cycle sequence (the numbers represent number of cycles): 4 ZnTe – 1 CdTe – 3 ZnTe – 2 CdTe – 2 ZnTe – 3 CdTe – 1 ZnTe – 4 CdTe; for a total of 20 cycles. A thicker sample B was grown starting with 7 cycles of ZnTe and following a sequence 7 (ZnTe) - 1(CdTe) - 6 (ZnTe) - 2(CdTe)…; for a total of 56 cycles. These two samples were grown onto GaAs substrates. For transmittance measurements a sample C was grown with the same conditions of sample B but using a glass/ITO substrate. Moreover, samples of pure ZnTe and CdTe with 20 cycles and the same growth temperature were also grown for comparison, both in GaAs and in transparent glass/ITO substrate. Prior to the growth, the substrates were degreased in acetone and alcohol; GaAs (100) substrates were subjected to a further chemical etch in $5H_2SO_4:1H_2O_2:1H_2O$ solution for 3 min respectively, followed by HF (48 %) for 20 s. In all cases the exposure times to the sources were 20 s and the purge times 10 s for a total cycle duration of 60 s.

The idea of these experiments is that Zn/Cd inter-diffusion (ID) must play a role in transforming the successive ZnTe and CdTe stacked layers on a relatively smooth ternary alloy with graded composition. If there would be excessive ID, composition gradients would cancel out and consequently a ternary alloy with a constant composition will be obtained after the growth process. On the other hand, if ID were insignificant, instead a GC layer, alternated CdTe and ZnTe films with different thickness will be obtained. Then, for a GC film to be formed, a moderate Zn/Cd inter-diffusion should be present. The criterion we followed as a guide to estimate the proper growth conditions was consider that the interface spreading due to ID, $2\sqrt{Dt}$ (with $D$ the Zn/Cd ID coefficient and $t$ the growth duration) must be of the same order of magnitude of the thickness grown per cycle. With this consideration we expect that local diffusion will be enough for smoothing the ZnTe/CdTe interfaces without forming a constant composition across the whole sample. The growth rate of pure CdTe and ZnTe were calibrated in previous experiments to be 8 and 4 nm/cycle, the growth process duration for the samples grown with 20 cycles (at a rate of 1 min per cycle) was 20 min. With this time and the condition for the interface spreading to be 6 nm (the average of the growth rates of CdTe and ZnTe) we estimate the ID coefficient ($D$) as 1 x 10$^{-16}$ cm$^2$/s. According the temperature dependence data of $D$ for Cd/Zn ID in CdTe/ZnTe heterostructures [19] this coefficient corresponds to a temperature of 400 $^0$C, which was used in the growth experiments.

2.2. Structural characterization and composition profiles

To determine the structure of the samples XRD θ–2θ scans were performed using a Siemens D-5000 powder diffractometer with the wavelength (λ) of Cu K$_{\alpha 1}$ radiation. Reciprocal space maps were obtained for the samples grown onto GaAs substrates by means of a high resolution PANalytical XPert PRO MRD diffractometer, using a Pixcel solid state detector at $0.05^0$ omega increments.

SIMS and XPS were used to obtain a qualitative or semi-qualitative evaluation of the composition profiles. SIMS measurements were made using a Cameca 6F IMS spectrometer. In order to reduce the matrix effect a $CsM^+$ mode was used. XPS depth profile analysis of the samples was performed on a VG-ESCALAB 220i-XL system (base pressure of $2 \times 10^{-10}$ mbar) equipped with photoelectron spectroscopy (XPS) and ion sputter gun for sample etching. The Al K$\alpha$ line (1486.6eV) was used as X-ray source. As XPS is a surface sensitive technique, giving chemical information of the last atomic layers of the film, a depth profile of the sample in terms of photoemission line intensities was obtained by combining a sequence of etching cycles interleaved with photoemission measurements. Argon ion sputtering cycles were performed with ion beam energy of 2keV. The As2p, Ga2p, Cd3d, Te3d, Zn2p lines were monitored as function of etching cycles.

Optical transmission measurements were performed using a double beam Hitachi UV-VIS 150-20 photo-spectrometer equipped with an integrating sphere. Normal incidence was used in the range 200-900 nm.

RBS measurements were done in order to obtain the composition profile and to determine the stoichiometry and the thickness of the samples. A 7 MeV alpha beam provided by a 3 MV Tandetron accelerator was employed. In this way we were able to obtain a good mass resolution in order to be able to identify the Te and Cd signals. The resolution of the detector and associated electronics was better than 13 keV.

3. Results and discussion

Fig. 1 shows the diffractograms for samples A and B. Also, diffractograms for CdTe and ZnTe samples grown at the same temperature and with the same number of cycles that sample A are shown for comparison. As can be seen in this figure, for samples A and B only the peaks corresponding to the (200) and (400) reflections are observed. Since the used GaAs substrate in the present experiments was (100) oriented, this is an indication of the epitaxial relation between the layer and the substrate. The same characteristic can be seen in the sample of pure ZnTe. That means that when the growth begin with ZnTe, the [100] crystalline direction is preserved also in mixed films. Instead, polycrystalline growth is induced as observed in the diffractogram of the pure CdTe sample grown at the same conditions. This is a consequence of the larger lattice mismatch of the CdTe/GaAs interface with respect the ZnTe/GaAs one.

Moreover, it can be also noted from Fig. 1, that samples A and B show much wider diffraction peaks as compared with that of the pure binary compounds. In fact, the width of these peaks cover a huge part of the complete $2\theta$ range from the pure CdTe to the ZnTe expected peak positions. So wide peaks can only be due to a variable lattice constant value originated in varying composition along the sample.

To better understand the structure of these samples, reciprocal space maps of the (004) region, including the substrate were taken. In Fig. 2 such maps are shown for samples A and B. The widths of the peaks along $\omega$ are 1.99 and 1.45 ° for layers A and B, respectively. They should be compared with 0.0139° for the substrate. This indicates that the growth is epitaxial but with high mosaicity, which is slightly smaller for the thicker B sample. This spread in $\omega$ is also larger than that typically observed in our ZnTe samples [17] whose values have been found to be about half of those reported above for GC samples. Along $2\theta$ the FWHM values are 0.112 and around 1° for the substrate and the layer, respectively. As indicated above, this higher value of the films is caused by the composition grading.

Fig. 3 shows the absorbance spectra for sample C together with two reference samples of CdTe and ZnTe, all of them grown onto transparent glass/ITO substrates. As can be observed, pure compounds present a sharp absorption edge at the corresponding band gaps. For sample C (equivalent to sample B), instead, the increase of the absorbance extends in a relatively large range as expected for a graded band gap material. This extended absorption is considered to be one of the advantages of GC layers in solar cell devices.

Fig. 4 shows a qualitative composition profile obtained by SIMS for sample A. The raster was 250 µm², while the detection area was 60 µm². From this figure it is possible to assess the graded composition of Zn and Cd: Cd increasing (and Zn decreasing) monotonically toward the surface as expected for this sample. At the same time, Te composition remains constant as corresponds to a $Cd_xZn_{1-x}Te$ stoichiometric pseudo-binary alloy.

XPS profile for the same sample is shown in Fig. 5 (the depth top scale was calibrated according the thickness calculated from RBS measurements as can be seen below). Also in this profile, it can be observed an increase (decrease) of Cd (Zn) toward the surface.

RBS technique was used to calculate the thickness and the composition profiles of GC samples. It is well known that RBS technique is sensitive not only to the chemical composition and stoichiometry but also to the depth of the scatter atom in the film. For this reason, it is a quite suitable technique for determining composition profiles.

In Fig. 6 the RBS spectra of samples A and B are shown. In order to obtain the corresponding compositions we have used the SIMNRA computer simulation code [20]. To this end, each layer was supposed to be composed by several sub-layers with different Zn, Te and Cd concentrations and thickness. It was observed, in the simulation of the RBS spectra a stoichiometric (Zn+Cd)/Te : 1/1. Then it was assumed that the films consist of $Cd_xZn_{1-x}Te$ pseudo-binary alloy in agreement with XRD patterns which did not show the presence of other phases different to $Cd_xZn_{1-x}Te$. The simulations are also shown in Fig. 6, including the individual contribution of each element. As can be seen from this figure, the simulation reproduces reasonably well both RBS spectra.

The obtained compositional profiles are displayed in Fig. 7. Both samples show relatively smooth decrease of Cd concentration from the surface to the interface with the substrate. This is expected taking into account the cycles sequences of these samples. Near the interface with the substrate, the ternary alloy has an intermediate composition. This fact can be explained considering that the region near the interface with the substrate was subjected by a longest time to the growth temperature. Then, the ID was enhanced there, this leading to a higher Zn/Cd mixing.

It is also observed that according the composition evaluation of the XPS and RBS profiles, the overall composition of Cd is higher than that of Zn. This is expected because the growth rate (growth thickness for cycle) is larger for Cd, and comprehensively, the number of cycles of Cd and Zn are the same. In fact, according to the cycle sequence for sample A, they were grown in total, 10 cycles of CdTe and 10 cycles of ZnTe, they correspond with a total amount of CdTe and ZnTe of 80 and 40 nm respectively. It leads to an average composition of CdTe and ZnTe of 66 and 33 %, respectively, which agree rather well with the profile obtained in the RBS spectra.

4. Conclusion

$Cd_xZn_{1-x}Te$ films with graded composition were obtained by the simple Isothermal Close Space Sublimation (ICSS) growth technique. For the preparation of the samples, CdTe and ZnTe layers were grown successively in such a way that the thickness of both layers (controlled by the number of cycles) was modified along the structure to produce an increase of the average concentration of CdTe. The growth temperature was selected to ensure a reasonable inter-diffusion among the layers such that the succession of layers was turned into a relatively smooth graded composition film.

Wide diffraction peaks in XRD patterns, SIMS and XPS profiles indicated the variable composition in the samples. Quantitative composition profiles were obtained by simulation of the RBS spectra. For the two samples considered here graded composition were obtained. At the surface, the composition was near to CdTe for both samples, while at the interface with the substrate the material presented an intermediate composition due to enhanced inter-diffusion in this region that was kept at the growth temperature during the longest time.

This work demonstrates the possibility of growing graded composition layers using a simple and cost-effective procedure. The composition profile can be, in principle, tuned by choosing the growth temperature, as well as the thickness and sequence of the binary layers. The procedure presented here could be interesting for solar cells technology community, for incorporating $Cd_xZn_{1-x}Te$ layer of variable composition in CdS/CdTe solar cells.

Acknowledgements. This work was partially supported by the CAPES-MES CUBA project 121/11 and IPN project 20120917. O. de Melo thanks CAPES visiting professor fellowship (BEX 14582139). The authors thank A. Tavira for reciprocal maps measurements and S. de Roux for technical assistance. The authors thank the Brazilian funding agencies CAPES, CNPq and FAPEMIG.

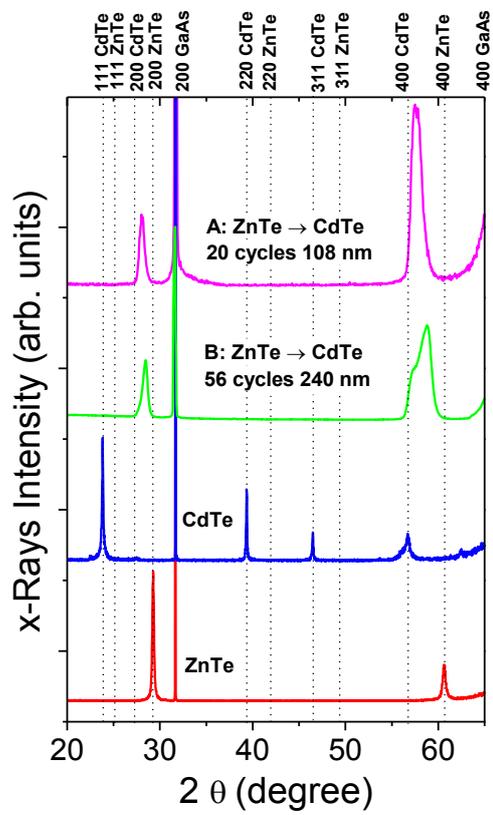

Fig. 1. De Melo et al. (color on the Web and in black-and-white in print)

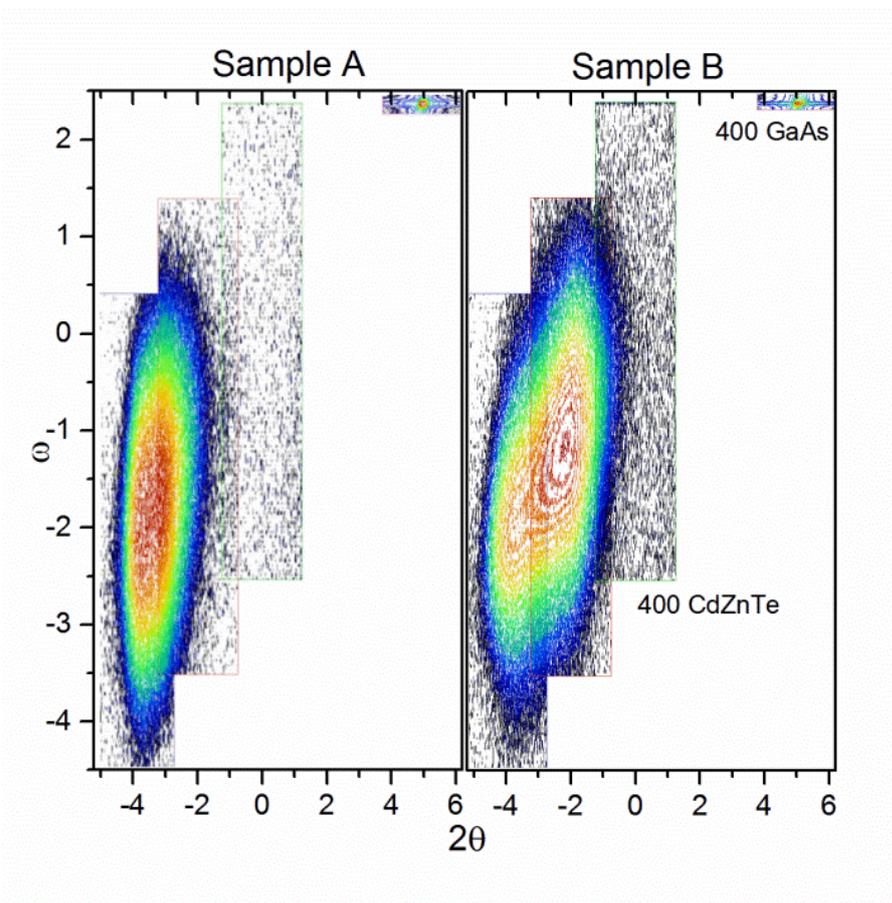

Fig. 2. De Melo et al. (color on the Web and in black-and-white in print)

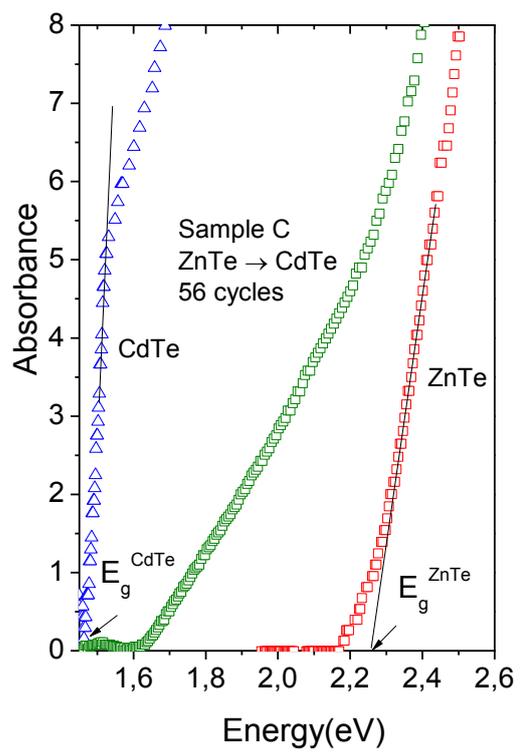

Fig. 3. de Melo et al. (color on the Web and in black-and-white in print)

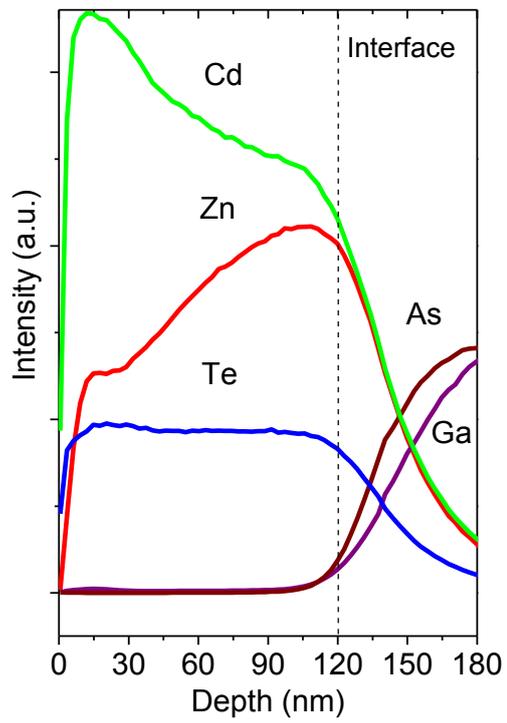

Fig. 4 de Melo et al. (color on the Web and in black-and-white in print)

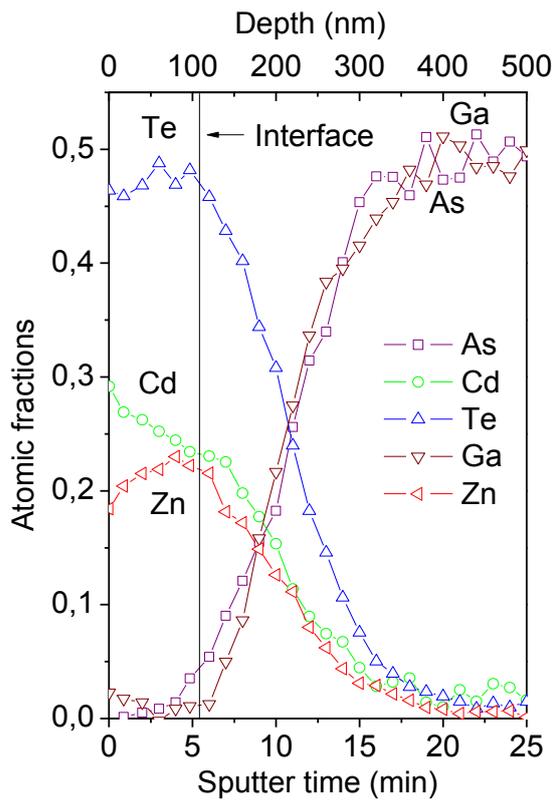

Fig. 5. de Melo et al. (color on the Web and in black-and-white in print)

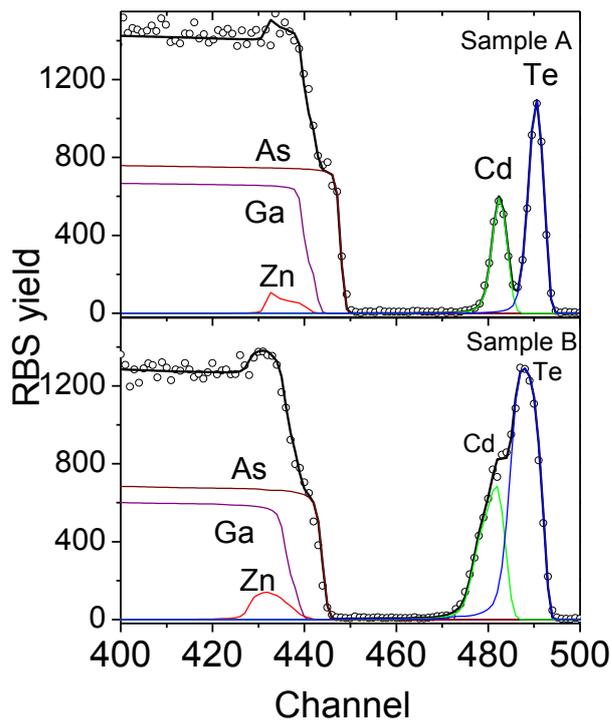

Fig. 6. de Melo et al. (color on the Web and in black-and-white in print)

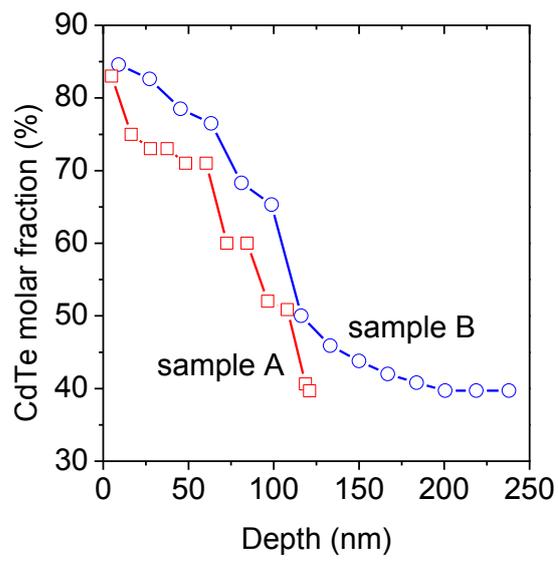

Fig. 7. de Melo et al. (color on the Web and in black-and-white in print)

Figure captions.

Fig. 1. XRD Diffractograms for samples A and B grown with cycles sequences explained in the text. Diffractograms of pure binary compounds films grown at the same temperature and with the same number of cycles of samples A are also shown.

Fig. 2. Reciprocal space maps for samples A and B in the $\omega$-$2\theta$ representation.

Fig. 3. Absorbance spectra for sample C, grown with the same conditions of sample B but using a transparent substrate. The spectra of reference films of pure CdTe and ZnTe are also shown.

Fig. 4. SIMS profiles for sample A showing graded composition of Cd and Zn along the film thickness.

Fig. 5. XPS depth profiles for sample A showing graded composition of Cd and Zn along the film thickness.

Fig. 6. RBS spectra for samples A and B (open dots). The simulation of the spectra and the contribution of the different elements are represented by thick and thin solid lines, respectively.

Fig. 7. CdTe molar fraction (expressed in per cent) profiles obtained by the simulation of the RBS spectra for samples A and B.